\documentclass[twocolumn,aps,prl,groupedaddress,reprint]{revtex4}
\usepackage{graphicx}
\usepackage{color}
\usepackage{amsmath}
\usepackage{marvosym}

\newcommand{\be}{\begin{equation}}
\newcommand{\ee}{\end{equation}}
\newcommand{\bea}{\begin{eqnarray}}
\newcommand{\eea}{\end{eqnarray}}

\newcommand{\la}{\langle}
\newcommand{\ra}{\rangle}

\renewcommand{\vec}[1]{{\bf #1}}
\renewcommand{\epsilon}{\varepsilon}

\begin{document}

\title{Linear magnetoresistance in metals: guiding center diffusion in a \\ smooth random potential} 
\author{Justin C. W. Song$^{1,2}$, Gil Refael$^{1,2}$, and Patrick A. Lee$^{3}$}
\affiliation{$^1$ Walter Burke Institute for Theoretical Physics and Institute for Quantum Information and Matter, California Institute of Technology, Pasadena, CA 91125 USA}
\affiliation{$^2$ Department of Physics, California Institute of Technology, Pasadena, CA 91125 USA}
\affiliation{$^3$ Department of Physics, Massachusetts Institute of Technology, Cambridge, MA 02139 USA}
\begin{abstract}
We predict that guiding center (GC) diffusion yields a
linear and non-saturating (transverse) magnetoresistance in 3D metals. Our theory is semi-classical and applies in the regime where the transport time is much greater than the cyclotron period, and for weak disorder potentials which are slowly varying on a length scale much greater than the cyclotron radius. Under these conditions, orbits with small momenta along magnetic field $B$ are
squeezed and dominate the transverse conductivity. When disorder potentials are stronger than the Debye frequency, linear magnetoresistance is predicted to  survive up to  room temperature and beyond. We argue that magnetoresistance from GC diffusion explains the recently observed giant linear magnetoresistance in 3D Dirac materials.
\end{abstract} 

\maketitle

Magnetoresistance provides a powerful means with which to probe the scattering history of particles in a magnetic field. Departure from the conventional paradigm - quadratic magneto-resistance at low fields, saturating at high fields~\cite{olsen} - signals anomalous particle scattering behavior. One particularly appealing regime is non-saturating and linear magnetoresistance (LMR), which has a long standing history~\cite{xu,rosenbaum1,kapitza,yang} given its potentially disruptive technological impact~\cite{techimpact}. 

Very few theories predict LMR in a closed single component Fermi surface. A well known example is Ref.~\cite{abrikosov}, which showed that Dirac metals in the extreme quantum limit (when only the $n=0$ Landau level is occupied) exhibit LMR in the presence of screened Coulomb impurities. Another mechanism yielding quasi-linear MR arises from inhomogeneity~\cite{herring,dreizendykne,parishlittlewood}. However, significant LMR in these requires strong inhomogeneity~\cite{parishlittlewood}. 
Contemporary proposals that extend the above treatments, have also found LMR under similar requirements~\cite{mirlin,shaffique}.

Recently, giant LMR that lie outside the above two paradigms [see (i) and (ii) below] was reported in the newly discovered class of three-dimensional Dirac materials (3DDM)~\cite{ando,coldea,ongcd3as2,ongNa3Bi,daiTaAs2}. LMR in 3DDM exhibit puzzling features including (i) its occurrence when multiple Landau levels are occupied far from the extreme quantum limit, and (ii) arising in weakly disordered, high mobility samples. 
Further, LMR manifests consistently over a variety of 3DDM experiments, including in TiBiSSe~\cite{ando}, Cd$_3$As$_2$ \cite{coldea,ongcd3as2}, Na$_3$Bi\cite{ongNa3Bi}, and TaAs~\cite{daiTaAs2}, where chemical potential $\mu$  typically lies 0.1 eV above the Dirac point, hinting at a single underlying explanation. 

\begin{figure}
\includegraphics[scale=0.155]{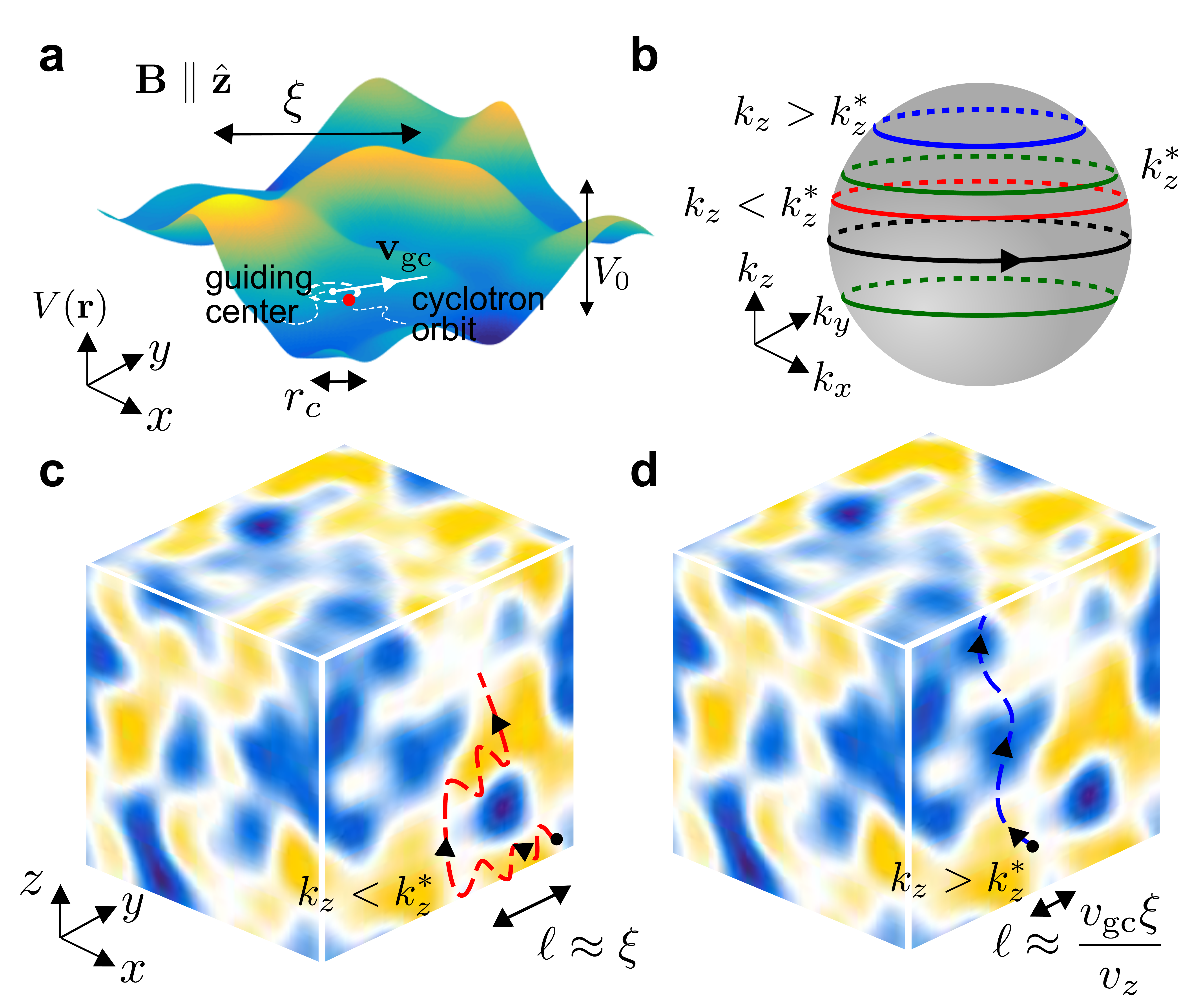}
\caption{a) Magnetoresistance can be dominated by guiding center (GC) motion when disorder correlation lengths $\xi \gg r_c$. Here $r_c$ is the cyclotron radius. This regime is characterized by slow GC motion, $\vec v_{\rm gc}  = \nabla_\vec{r} V(\vec r)\times \hat{\vec z}/B$, accompanied by fast cyclotron orbits, $\mathbf{v}_{\rm cycl}$. GC diffusion in this environment givers rise to LMR. b) Electrons perform closed orbits of the Fermi surface, with GC motion classified into two types, $k_z< k_z^*$ (red), and $k_z > k_z^*$ (blue); critical $k_z^*$ (green). c) For $k_z< k_z^*$, electrons are squeezed in $z$ yielding mean free paths $\ell \approx \xi$ and in-plane $D_{xx} \sim v_{\rm gc} \xi \propto 1/B $. d) In contrast, $k_z > k_z^*$ electrons exhibit unconstrained $z$ motion yielding in-plane $D_{xx} \sim v_{\rm gc}^2 \xi/v_z \propto 1/B^2$ (see text).} 
\end{figure}

Here we propose a semi-classical mechanism for 
LMR in metals, wherein charge transport is dominated by guiding center (GC) motion. Importantly, this mechanism naturally gives giant LMR under (i) and (ii) above, explaining the puzzling behavior~\cite{ando,coldea,ongcd3as2,ongNa3Bi,daiTaAs2}. The main requirement is that the disorder potential is smoothly varying on a scale, $\xi$, which is large as compared to the cyclotron radius, $r_c$. The main features of GC magnetoresistance are exposed by writing 
the transverse resistivity as 
\be
\rho_{xx} = \frac{\sigma_{xx}}{\sigma_{xx}^2 + \sigma_{xy}^2} = \frac{\mathcal{G}}{\sigma_{xy}}, \quad \mathcal{G} = \frac{{\rm tan} \theta_H}{1+ [{\rm tan} \theta_H]^2},
\label{eq:rhoidentity}
\ee
where $\sigma_{xx}$ and $\sigma_{xy}$ are the transverse ($x$-$y$ plane) conductivity and Hall conductivities respectively, and ${\rm tan} \theta_H = \sigma_{xy}/\sigma_{xx}$ is the Hall angle. Using the familiar $\sigma_{xy} = ne/B$ with $n$ the density, and $e$ the carrier charge, we have
\be
\rho_{xx} = \frac{B\mathcal{G}}{ne}.
\label{eq:linearMR}
\ee
As we argue below, in the regime of $\omega_c \tau_{\rm tr} \gg 1$ and $\xi \gg r_c$, 
GC diffusion gives a Hall angle, and therefore $\mathcal{G}$, that is {\it independent} of magnetic field magnitude, leading to LMR in Eq.~(\ref{eq:linearMR}).  Here $\omega_c$ is the cyclotron frequency, $\tau_{\rm tr}$ is the transport time, and the magnetic field $\vec B = B \hat{\vec z}$.

Guiding center magnetoresistance can be understood as follows. In semi-classically large $B$ fields ($\omega_c \tau_{\rm tr} \gg 1$), electrons exhibit in-plane trajectories $\vec r_\perp (t)$ characterized by slow guiding center motion $\vec R (t)$ accompanied by fast cyclotron orbits $\vec r_{\rm cycl} (t)$. The latter, characterized by $r_c$, depends on intrinsic material properties and $B$; whereas the former depends on the potential profile sampled by the electron over one cycle which can yield unusual trajectories~\cite{beenaker}. A unique situation arises for slowly varying disorder potentials $V(\vec r)$. In this regime $\xi \gg r_c$ (see Fig. 1a), electron trajectories are dominated by guiding center motion which follows the local disorder landscape at $R$, with velocity $\mathbf{v}_{\rm gc} = [\nabla_{\vec{R}} V(\vec R)] \times \hat{\mathbf{z}}/B$. 

Guiding center diffusion is characterized by diffusion constant $D^{\rm gc}_{xx} \sim v_{\rm gc}^2 \tau$. The central question is: {\it what is} $\tau$? First, it is important to note this picture is not valid in strictly 2D, because GCs form closed orbits along equipotential lines. Hence it is crucial to include motion in the $z$ direction which restores diffusive motion. There are two classes of electron motion depending on their $k_z$ value with respect to $k_z^*$ (Fig. 1b). For $k_z > k_z^*$, electrons possess
kinetic energy in the $z$ direction exceeding the typical potential fluctuation. As a result, the electron moves freely across many potential fluctuations shown in Fig. 1d. Within time $\tau_>\approx \xi/v_z$, the GC senses a different local electric field and changes direction. As a result, $D^{\rm gc}_{xx}(k_z>k_z^*)  \sim  v_{\rm gc}^2 \tau_> \propto 1/B^2$. Using $\sigma_{xy}=ne/B$ and Eq.~(\ref{eq:rhoidentity}), we recover the standard saturated magnetoresistance. 

On the other hand, electrons with $k_z<k_z^*$ are typically 
squeezed by a local potential barrier. As shown in Fig. 1c, they must travel sideways by a distance $\xi$ to get around the barrier. In this case, $\tau_< \approx \xi/v_{\rm gc}$ and $D^{\rm gc}_{xx}\sim v_{\rm gc} \xi \sim 1/B$; LMR follows immediately from Eq.~(\ref{eq:rhoidentity}) and (\ref{eq:linearMR}). Paradoxically for large $B$, it is the electrons squeezed in the $z$ direction which dominate transport in the $x$-$y$ plane, leading to LMR. 
The importance of squeezed electrons was pointed out by Ref.~\cite{polyakov1986,murzinA} in the context of magneto-transport of Boltzmann gases. We adapt their reasoning to the case of a degenerate Fermi sea and present a more quantitative treatment below.

We emphasize that our mechanism for LMR does not have anything to do with the Dirac spectrum per se. Nevertheless, 3D Dirac semi-metals provide an ideal venue that satisfy the conditions required for LMR. First, the high mobility of 3DDM ($\eta \gtrsim {\rm few} \times \, 10^4 {\rm cm}^{2}/ {\rm Vs} $) allows the regime where electrons undergoes many cyclotron orbits before scattering, $\omega_c \tau_{\rm tr}  = B \eta \gg 1$, to be achieved at relatively low magnetic fields.  Second, the relatively small chemical potential $\mu \sim 100 \, {\rm meV}$ {\it but} large Fermi velocity $v_F \sim 1 - 10 \times 10^8 \, {\rm cm}/{\rm s}$~\cite{coldea}, give small cyclotron radius $r_c \sim \mu/ ev_F B =  10 - 100 \, {\rm nm}$ even at $1 \, {\rm T}$. Third, large dielectric constants of $\kappa \sim 40$~\cite{jaygerin,skinner} effectively screen Coulomb impurities to yield weak and slowly varying disorder potentials, with large correlation lengths, $\xi \sim 20 - 60 \, {\rm nm}$~\cite{skinner}. As a result, $\xi \gg r_c$ at relatively low $B \gtrsim 1 \, {\rm T}$, allowing GC motion to dominate the $x$-$y$ plane (transverse) magnetoresistance.

We note parenthetically that our regime of interest is distinct from multi-component systems e.g., charge compensated systems which can be ultra-sensitive to magnetic field~\cite{pkbook,narozhny}. 
Instead we are interested in LMR in 3DDM~\cite{ando,coldea,ongcd3as2,ongNa3Bi,daiTaAs2} which were observed in the metallic regime with carriers in a single band~\cite{ongcd3as2}.

We begin by considering the diffusive motion of charged particles in a magnetic field, $\mathbf{B} = B \hat{\vec z}$, and a slowly varying and weak disorder potential $V(\vec r)$. While formally interested in 3DDM, our analysis below is general; we will only specify 3DDM as needed to compare to recent experiments. Disorder is characterized by $\la V(\vec r) V(\vec r') \ra = V_0^2 \mathcal{F} (|\vec r - \vec r'|/\xi)$, where $\la \mathcal{O} \ra$ denotes disorder averaging, and $\xi$ the correlation length.  $\mathcal{F}$ is a dimensionless function that vanishes for $|\vec r - \vec r'| \gg \xi$. Lastly, we will be interested in weak disorder strength $eV_0 < \mu$ seen in 3DDM experiments~\cite{ando,coldea,ongcd3as2,ongNa3Bi,daiTaAs2}, and recent estimates~\cite{skinner}. 

{\it Equations of motion} - The motion of particles on the Fermi surface with chemical potential $\mu$ can be described by the semi-classical equations of motion 
\begin{subequations}
\begin{eqnarray}
&& m\dot{\vec v}_\perp = -e \nabla_{\vec r}V(\vec r) + e \vec v_\perp  \times \vec B,  \\
&&m\dot{v}_z = -e \partial_z V (\vec r),
\end{eqnarray}
\label{eq:eom}
\end{subequations}
where $m$ is the cyclotron mass, $\vec v_{\perp} (k_z,\mu) = (v_x,v_y)$, and $v_z (k_z,\mu)$ are velocities transverse to the magnetic field and along the magnetic field respectively. We note that throughout our analysis below, these quantities depend on momentum along the field, $k_z$, and $\mu$. For e.g., velocity is captured via group velocity $\vec v_{\vec k} = \hbar^{-1}\partial \epsilon_{\vec k}/ \partial \vec k$ so that the $x$-$y$ plane speed for Dirac particles is $|\vec v_\perp| = v_0 = v_F \sqrt{1-k_z^2/k_F^2}$, and $m = \mu/ v_F^2$~\cite{novosolev}, where $v_F$ is Fermi velocity, and $k_F$ the Fermi wave vector; $\epsilon_{\vec k}$ is the particle dispersion. For brevity, we will drop explicit mention of $k_z$ dependence, bringing it up when necessary. We do not expect Berry phase related terms~\cite{sonspivak} to contribute to the transverse magnetoresistance behavior that we are interested in here.

The trajectories of charged particles, $\vec r(t)$, in crossed $\vec B$ and $V(\vec r)$ can be complex, since they involve transport processes spanning multiple time scales (e.g., cyclotron period, guiding center scattering time, and transport time). However, in semi-classically strong fields ($\omega_c \tau_{\rm tr} \gg 1$), and for a slowly varying potential so that correlation length is larger than cyclotron radius ($\xi \gg r_c$), its motion is conveniently captured via $\vec r (t)  = \vec R (t) + \vec r_{\rm cycl}(t)$. Here $\vec R(t)$ is the slow moving 3D guiding center coordinate, whereas $\vec r_{\rm cycl} (t)$ describes fast cyclotron motion lying in the $x$-$y$ plane. 

This reasoning yields the following ansatz for velocity in the $\vec r_\perp = (r_x, r_y)$ plane $\vec v_\perp$ as~\cite{beenaker}
\be
\tilde{v}_{\perp} (t) = v_0 e^{i\omega_c t} + \tilde v_{\rm gc}(t), \quad \tilde{v}_{\rm gc}  =  \frac{i \tilde{E} (\tilde{r}_\perp) }{B}, 
\label{eq:ansatz}
\ee
where  $\omega_c = eB/m$, and used complex notation $\mathcal{O}_x + i \mathcal{O}_y = \tilde{\mathcal{O}}$ for vectors in the $x$-$y$ plane.  
The latter part of Eq.~(\ref{eq:ansatz}) was obtained by substituting the ansatz into Eq.~(\ref{eq:eom}a) and setting $m \dot{v}_{\rm gc} = 0$ for slowly varying $V(\tilde r)$. Eq.~(\ref{eq:ansatz}) is valid for $| m \dot{v}_{\rm gc}| \ll  |e E(\tilde{r}_\perp) |$. Estimating $E \approx V_0/\xi$, we obtain the condition
\be
 \xi^2 \gg  \frac{eV_0}{\omega_c^2 m} = r_c^2 \frac{eV_0}{\mu} , \quad {\rm where }\quad r_c  = \frac{v_0}{\omega_c}. 
\label{eq:condition}
 \ee 
Since we are interested in weak disorder $eV_0 < \mu$,  
the above condition is satisfied within our regime of validity, $\xi \gg r_c$. 

Motion in $z$ can be understood in the following way. First, we note that for $\xi \gg r_c$ and $\omega_c\tau_{\rm tr} \gg 1$, the potential the electron feels is determined by $\la \vec r  (t) \ra_{\rm 1 \, cycle}= {\vec R}(t)$. Next, for $ v_{\rm gc} \ll v_z$ electrons, the GC moves slowly in the $x$-$y$ plane as compared with $z$. As a result, integrating Eq.~(\ref{eq:eom}b) yields energy conservation
\be
\frac{m}{2} \Big\{ v^2_z [\vec R (t)] - v^2_z [\vec R (0)]\Big\} = -e \Big\{V[\vec R (t)] - V[\vec R (0)]\Big\},
\label{eq:condition2}
\ee
where we have set $\nabla_{\vec r_\perp}  V (\vec r)  \cdot \partial_t \vec r_\perp =0$. This is valid when $|\nabla_{\vec r_\perp}  V (\vec r)  \cdot \partial_t \vec r_\perp | \ll |v_z \partial_z V(\vec r)|$. Estimating $| \partial_t \vec r_\perp | \approx v_{\rm gc}$ and using disorder that is isotropic, yields the original condition $v_{\rm gc} \ll v_z$. 

{\it Guiding Center Transport -} The separation of time scales between slow GC motion, and fast cyclotron motion enables us to write the velocity correlator as
\be
\la v_\perp (t) v_\perp (0) \ra \approx \la v_{\rm gc} (t) v_{\rm gc} (0) \ra + v_0^2 e^{i\omega_c t - t/\tau_{\rm tr}},
\label{eq:vcorrelator}
\ee
where we have used a relaxation-time approximation in the last term to capture the Drude contribution to magnetotransport~\cite{beenaker}. 

Replacing $\vec r (t) $ with its average over one cycle as above, and using Eq.~(\ref{eq:ansatz}), we find GC diffusion, $D_{xx}^{\rm gc} = (1/2) \int_0^\infty \la v_{\rm gc} (t) v_{\rm gc} (0) \ra dt $, as
\begin{align}
D_{xx}^{\rm gc} & =   \int_0^\infty \la E[\vec R(t)] E[\vec R(0)]\ra dt/ (2B^2) = E_0^2  \tau/(2B^2), \nonumber \\ \tau & = \int_0^\infty dt \mathcal{F} (\Delta R / \xi),
\label{eq:dxxdefn}
\end{align}
where 
$\Delta R= | \vec R(t) - \vec R(0) |$, $E_0$ is the characteristic electric field strength of the disorder potential, and $\tau$ is the scattering time that is sensitive to the GC trajectory. 

We adopt a mean-field approach in estimating $\tau$.
Since $\mathcal{F} $ rapidly decays for $\Delta R > \xi$, $\tau$ is most sensitive to the way the GC moves in $\Delta R (t) < \xi$. As a result, we write $d\Delta R= v_{\rm av} dt$, with speed $v_{\rm av} = [\la v_{\rm gc}\ra_{\xi} ^2 + \la v_z\ra_{\xi}^2 ] ^{1/2}$ averaged over a single domain; here $\la \mathcal{O} \ra_\xi$ denotes averaging across a single domain. Changing variables $t \to \Delta R$, yields
\be
\tau \approx \frac{\xi \mathcal{A}}{\big[\la v_{\rm gc}\ra_{\xi} ^2 + \la v_z\ra_{\xi}^2\big] ^{1/2}}, \quad \mathcal{A} = \int_0^\infty dx \mathcal{F} (x) 
\label{eq:dgc-implicit}
\ee
where $\mathcal{A}$ is a number of order unity. Using gaussian correlations, $ \la V(x) V(0) \ra = V_0^2 \mathcal{F}(x) = V_0^2 e^{- x^2/\xi^2}$ we obtain $\mathcal{A} = \sqrt{\pi}/2$, and $E_0^2 \xi^2 = 6 V_0^2$.

Two distinct classes of GC trajectories can be discerned: (a) squeezed $z$-motion (Fig. 1c), and (b) unrestricted $z$-motion (Fig. 1d). Squeezing in class (a) arises from energy conservation in Eq.~(\ref{eq:condition2}):  
for particles with $m v_z^2/2 < V(\vec r)$, $z$-motion is constrained 
within a $V(\vec r)$ puddle. It escapes when GC 
diffuses out of the $V(\vec r)$ puddle (Fig. 1c).  
Squeezing yields $\la v_z \ra_\xi $ that {\it vanishes} and $v_{\rm av} \approx \la v_{\rm gc} \ra_\xi \approx E_0/B$. As a result, Eq.~(\ref{eq:dgc-implicit}) yields $\tau_< \approx \xi \mathcal{A}/ \la v_{\rm gc}\ra_{\xi} $ and 
\be
D_{xx}^{\rm gc} =   \frac{E_0\xi \mathcal{A}}{2 B},  \quad {\rm for} \quad v_z \lesssim (e2V_0/m)^{1/2} = v_*,
\label{eq:dxx}
\ee
corresponding to electrons in Fig. 1b with $k_z < k_z^*$; $k_z^*$ depends on the dispersion relation and $v_*$. For 3DDM, $\hbar k_z^* = (e2 V_0 \mu/ v_F^2)^{1/2}$. We note that Eq.~(\ref{eq:condition2}) can only be used for electrons with $v_z \gg v_{\rm gc}$, Eq.~(\ref{eq:condition2}). However, in the opposite limit $v_z \ll v_{\rm gc}$,  $\la v_z \ra_\xi^2$ is obviously smaller than $\la v_{\rm gc} \ra_\xi^2$ in Eq.~(\ref{eq:dgc-implicit}), allowing us to neglect the former's contribution, yielding $D_{xx}^{\rm gc}$ as in Eq.~(\ref{eq:dxx}). As a consistency check, we note that $v_* \gg v_{\rm gc}$ for our regime of validity $\xi \gg r_c$, $eV_0< \mu$~\footnote{Using Eq.~(\ref{eq:ansatz}),~(\ref{eq:dxx}), and $E_0 \approx  V_0/\xi$ yields $v_*/v_{\rm gc} \approx \sqrt{2} (\xi/r_c) \times (\mu/eV_0)^{1/2}$.}. Therefore $v_*$ determines the range of electrons that obey Eq.~(\ref{eq:dxx}).

In contrast to Eq.~(\ref{eq:dxx}), electrons with $v_z \gtrsim v_*$ do not have $z$-motion squeezed [case (b), see Fig. 1d]. As a result, GC samples many $V(\vec r)$ domains, with its $x$-$y$ plane velocity scrambled over times $\tau_> \sim \xi/v_z$, yielding an $x$-$y$ plane mean free path $ \ell \sim v_{\rm gc} \xi/v_z$. This is captured in Eq.~(\ref{eq:dgc-implicit}) whence $\la v_{\rm gc}\ra_\xi \ll \la v_z\ra_\xi$, giving $v_{\rm av} \approx \la v_z \ra_\xi \approx v_z$. As a result, Eq.~(\ref{eq:dgc-implicit}) yields $D_{xx}^{\rm gc} \propto 1/B^2$. Importantly, for sufficiently large $B$, $v_z > v_z^*$ electrons, while mobile in the $z$ direction, exhibit suppressed $x$-$y$ plane mobility as compared with $v_z < v_z^*$. As a result, class (a) trajectories dominates $x$-$y$ plane transport.

{\it Linear Magnetoresistance in 3DDM - } To illustrate the striking effects of GC diffusion we specialize to 3DDM. Using the Einstein relation,  
and Eq.~(\ref{eq:dxxdefn}-\ref{eq:dxx}), 
we obtain
\be
\sigma_{xx}^{\rm gc}  = e^2 \sum_{\vec k} D_{xx}^{\rm gc} (k_z) \delta (\epsilon_{\vec k} -\mu) =  \alpha \Big[\frac{1}{B} +\frac{\tilde{B}}{B^2}\Big], 
\label{eq:sigmaxx}
\ee
where $\epsilon_{\vec k}$ is the electron energy, $\alpha/(e^2  \nu_{\rm 2D}) \approx E_0 \xi \mathcal{A} k_z^*/(4\pi)$, and $\tilde{B} \approx (E_0/v_*) \times {\rm ln} (k_F/k_z^*)$. Here we have used the 2D density of states for a $k_z$ slice in 3DDM as 
$\nu_{\rm 2D}(\mu) = \sum_{k_x,k_y} \delta (\epsilon_{\vec k} -\mu) =\mu/2\pi \hbar^2 v_F^2$, and $D_{xx}^{\rm gc}(k_z)$ obtained from the two trajectory classes (a) and (b). We note that the first term dominates over the second when $B> \tilde{B}$. It is useful to re-write this condition as $r_c < \xi \mathcal{K}$, where $\mathcal{K} = \sqrt{2} (\mu/eV_0)^{1/2} \times ( {\rm ln} [k_F/k_z^*])^{-1} \gtrsim 1$, since we are interested in $\mu> eV_0$. Hence, the first term always dominates in our regime of validity, $\xi \gg r_c$. 

In the same way, the second term in Eq.~(\ref{eq:vcorrelator}) yields the usual expressions 
\be
\sigma_{xx}^{\rm cycl} = \sum_{k_z} \zeta \tau_{\rm tr}, \quad  \sigma_{xy} = \sum_{k_z} \zeta \omega_c \tau_{\rm tr}^2 \approx \frac {ne}{B}, 
\label{eq:cyclotronsigma}
\ee
where $\zeta = e^2 \nu_{\rm 2D}(\mu) v_0^2/[2(1+ \omega_c^2 \tau_{\rm tr}^2)]$; we have taken $\omega_c\tau_{\rm tr} \gg 1$ limit in the last expression. Since $\sigma_{xx}^{\rm cycl} \propto 1/B^2$, for sufficiently large fields it provides a negligible contribution to $\sigma_{xx}$ as compared with Eq.~(\ref{eq:sigmaxx}). 

An important diagnostic of magnetotransport is the Hall angle, ${\rm tan} \theta_H = \rho_{xy}/\rho_{xx} = \sigma_{xy}/\sigma_{xx}$. Using Eq.~(\ref{eq:sigmaxx}) and writing $\sigma_{xx} = \sigma_{xx}^{\rm gc}$ (neglecting $\sigma_{xx}^{\rm cycl}$ since $\omega_c \tau_{\rm tr} \gg 1$), we obtain a $B$-field 
magnitude independent 
\be
{\rm tan} \theta_H = \frac{2}{\sqrt{27\pi}} \Big(\frac{\mu}{eV_0}\Big)^{3/2}, 
\label{eq:hallangle}
\ee
where we have used $n = \mu^3/6\pi^2 \hbar^3 v_F^3$ for a single fermion flavor in 3DDM, and gaussian correlated $\la V(\vec r) V(\vec r') \ra$. We note that the Hall angle changes sign when $B$ field flips sign. Interestingly, the Hall angle can be tuned by $V_0$ and $\mu$. Estimating $V_0 \approx 20 \, {\rm mV}$ in 3DDM~\cite{skinner}, and $\mu \sim 0.1 \, {\rm eV}$~\cite{ando,coldea,ongcd3as2,ongNa3Bi,daiTaAs2}, we obtain ${\rm tan} \theta_H \approx 2.4$.

We note that tunable Hall angle [Eq.~(\ref{eq:hallangle})] controls $\mathcal{G}$. Indeed, $\mathcal{G}$ is a non-monotonic function of Hall angle (and hence it depends on $\mu/eV_0$), reaching a peak when Hall angle becomes unity.  

To summarize, we find that under the conditions $\omega_c\tau_{\rm tr} \gg 1$, $\xi \gg r_c$, $\mu > eV_0$, $\mathcal{G}$ is independent of $B$ field magnitude, yielding LMR according to Eq.~(\ref{eq:linearMR}). Of the first two conditions, $\xi \gg r_c$ can be expressed as $B>B_c =  m v_F/(e\xi)$ which is a more stringent condition than $\omega_c \tau_{\rm tr} \gg 1$. This is seen by estimating $\tau_{\rm tr}$ using the Born approximation, giving $B_c$ that exceeds the field marking the onset of $\omega_c\tau_{\rm tr} > 1$ by a factor $\sim (\mu/ eV_0)^2$. Hence, we predict LMR as long as $B>B_c$. 

An important figure of merit for magnetoresistance is the ratio ${\rm MR} = \big[\rho_{xx} (B) - \rho_{xx} (0)\big]/ \rho_{xx}(0)$. Using Eq.~(\ref{eq:linearMR}), and noting the mobility $\eta = \sigma_{xx} (0) /ne$, we obtain 
\be
{\rm MR} = \frac{\rho_{xx}(B) - \rho_{xx} (0)}{\rho_{xx}(0)} \approx \Big( \frac{\eta [ {\rm cm}^{2}/ {\rm Vs}]}{10^4}\Big)\times\, B [{\rm T}] \times  \mathcal{G}. 
\label{eq:MR}
\ee
For typical $\eta \approx 1- 20 \times 10^{4} \,{\rm cm}^{2}/ {\rm Vs}$ in 3DDM, Eq.~(\ref{eq:MR}) yields giant ${\rm MR} \approx 5 - 100 $ at $B= 10\, {\rm T}$. Here we have used maximal $\mathcal{G}= 1/2$. For fixed $\mathcal{G}$, Eq.~(\ref{eq:MR}) is consistent with Kohler's rule~\cite{normalmetals} since it scales with mobility. This mirrors the experiments where MR scaled with temperature dependent mobility over wide range of $B$~\cite{coldea}. 

Another intriguing feature of LMR is an unconventional Hall resistivity at large fields. Conventionally, $\rho_{xy}^{(0)} = B/ne$ at large fields and is used to determine the density of carriers. However, since $\sigma_{xx} \propto 1/B$ above, we obtain a Hall resistivity $\rho_{xy} = \sigma_{xy} / (\sigma_{xy}^2 + \sigma_{xx}^2) = \rho_{xy}^{(0)} \mathcal{G}_2$ that differs from the conventional case by a factor $\mathcal{G}_2 = {\rm tan} \theta_{\rm H} \times \mathcal{G}$. Since $ \mathcal{G}_2 \leq 1 $, small Hall angles, can lead to unconventional $\rho_{xy}$. Indeed, in these situations, determining density through Hall measurements via $n = B/(\rho_{xy} e)$ may overestimate the density. 

We note that for other dispersions, our treatment above follows through yielding LMR under the same conditions of  $\omega_c\tau_{\rm tr} \gg 1$, $\xi \gg r_c$, $\mu > eV_0$. However, features e.g., $k_z^*$, $\alpha$ and Hall angle are altered appropriately. 

{\it Inelastic scattering - } Energy relaxation through inelastic scattering (e.g. through phonon scattering) in the $z$-direction can drastically affect GC motion by mixing squeezed $z$ with unconstrained $z$ trajectories.  Absorption of phonons with $\hbar \omega \gtrsim V_0 - \frac{m}{2}v_z^2$ relaxes the energy constraint [Eq.~(\ref{eq:condition2})], allowing $v_z < v_*$ electrons to jump out of $V(\vec r)$ troughs. If $\tau_< \approx \xi/v_{\rm gc} \gtrsim \tau_{\rm in} (\epsilon = V_0 - \frac{m}{2}v_z^2)$, these electrons exhibit $D_{xx} (k_z) \propto 1/B^2$. Here $\tau_{\rm in} (\epsilon)$ is the time for an electron to absorb energy $\epsilon$. Suppressed at low T, phonon-assisted escape leads to LMR degradation when $k_B T \gtrsim V_0$.

However, when typical $V_0 \gg \hbar \omega_{\rm D}$, phonon-assisted escape becomes difficult even at high temperatures since the maximum energy that can be absorbed from phonons is $\hbar \omega_{\rm D} \ll V_0$; $\omega_D$ is the Debye frequency. As a result, in this regime LMR is stable even at high temperatures and large fields, as recently observed in Cd$_3$As$_2$~\cite{coldea,ongcd3as2} where LMR persisted at 300 K and high fields. 

While detrimental to LMR from GC diffusion in 3D, inelastic scattering
can have the opposite effect in 2D. 
Conventionally in 2D, GCs form closed orbits along equipotential lines of a disorder potential yielding localization behavior~\cite{prangeandgirvin}. However, for inelastic phonon scattering which is not so strong to entirely disrupt the GC motion, but strong enough ($\xi/v_{\rm gc} \gg \tau_{\rm in}$) to induce switching between adjacent equipotentials~\cite{zhao1993}, the GC trajectories can become {\it open}, moving through multiple $V(\vec r)$ domains.
In this regime, we speculate $D_{xx}^{\rm 2D} \sim v_{\rm gc} \xi \propto 1/B$ 
as above. Interestingly, 2D semi-classical regime LMR was reported previously~\cite{renard}.

Semi-classical GC diffusion can conspire to produce LMR in metals. Importantly, the requirements for GC magnetoresistance are modest - 
arising in the semi-classical regime with multiple occupied Landau levels, and for weak and smooth disorder. Giant MR ratios, $B$-field magnitude independent Hall angles, $\mu$ and $V_0$ tunability, and stability at high temperatures, make GC diffusion and its magnetoresistance easy to identify in experiment. Indeed, these features bear striking resemblance to LMR measured recently in a variety of 3DDM~\cite{ando,coldea,ongcd3as2,ongNa3Bi,daiTaAs2}. Additionally, oscillatory motion of the GC trajectories along $B$ could have interesting, polarization-dependent, absorption signatures in the Terahertz regime.

\begin{acknowledgments}
We thank Adam Nahum and Brian Skinner for helpful discussions. JCWS acknowledges support from a Burke fellowship at Caltech. GR acknowledges support from the Packard Foundation and the Institute for Quantum Information and Matter (IQIM) an NSF funded physics frontier center, supported in part by the Moore Foundation. PAL acknowledges the support of the DOE under grant DE-FG01-03-ER46076, and the hospitality of the IQIM while this work was initiated. 
\end{acknowledgments}


\begin{thebibliography}{99}

\bibitem{olsen} J. L. Olsen, {\it Electron Transport in Metals}, Interscience, New York, (1962).

\bibitem{kapitza} P. L. Kapitza,  
Proc. R. Soc. London A {\bf 119}, 358-443 (1928). 




\bibitem{xu} Xu, R. et al., Nature {\bf 390}, 57-60 (1997);  A.Husmann, et al., Nature {\bf 417}, 421
(2002); M. Lee, T. F. Rosenbaum, M.-L. Saboungi, and H. S. Schnyders, 
Phys. Rev. Lett. {\bf 88}, 066602 (2002). 



\bibitem{yang} F. Y. Yang, et al., 
Science {\bf 284}, 1335-1337 (1999).
\bibitem{rosenbaum1} J. Hu, T. F. Rosenbaum, 
Nat. Mat. {\bf 7}, 697-700 (2008).



\bibitem{techimpact} Y.-A. Soh, and G. Aeppli,  
Nature {\bf 417}, 392-393 (2002). 


\bibitem{abrikosov} A. A. Abrikosov, 
Phys. Rev. B {\bf 58}, 2788 (1998).

\bibitem{herring} C. Herring, 
J. App. Phys., {\bf 31}, 1939 (1960). 
\bibitem{dreizendykne}  Y. A. Dreizin, A. M. Dykhne, 
Sov. Phys. JETP {\bf 36}, 127-136 (1973).

\bibitem{parishlittlewood} M. M. Parish and P. B. Littlewood, 
Nature {\bf 426}, 162-165 (2003); {\it Ibid}., Phys. Rev. B {\bf 72} 094417 (2005).





\bibitem{mirlin} J. Klier, I.V. Gornyi,and A.D. Mirlin, 
arXiv: 1507.03481 (2015). 
 \bibitem{shaffique} N. Ramakrishnan, M. Milletari, and S. Adam, 
 arXiv: 1501.03815 (2015).

\bibitem{ando} M. Novak, S. Sasaki, K. Segawa, and Y. Ando, 
Phys. Rev. B {\bf 91}, 041203(R) (2015).
\bibitem{ongcd3as2} T. Liang, Q. Gibson, M. N. Ali, M. Liu, R. J. Cava and N. P. Ong, 
Nat. Mat. {\bf 14}, 280-284 (2015). 
\bibitem{coldea} A. Narayanan, M. D. Watson, S. F. Blake, N. Bruyant, L. Drigo, Y. L. Chen, D. Prabhakaran, B. Yan, C. Felser, T. Kong, P. C. Canfield, and A. I. Coldea, 
Phys. Rev. Lett. {\bf 114}, 117201 (2015).
\bibitem{ongNa3Bi} J. Xiong, S. Kushwaha, J. Krizan, T. Liang, R. J. Cava, and N. P. Ong, 
arXiv: 1502.06266 (2015).
\bibitem{daiTaAs2}  X. Huang, L. Zhao, Y. Long, P. Wang, D. Chen, Z. Yang, H. Liang, M. Xue, H. Weng, Z. Fang, X. Dai and G. Chen, 
arXiv: 1503.01304 (2015).


\bibitem{beenaker} C.W.J.  Beenaker, 
Phys. Rev. Lett. {\bf 62}, 2020 (1989).
\bibitem{polyakov1986} D. G. Polyakov, Sov. Phys. JETP {\bf 63} 322 (1986).
\bibitem{murzinA} S. S. Murzin, N. I. Golovko,  Sov. Phys. JETP {\bf 54} 166 (1991); see the following for a review: S. S. Murzin, 
Physics Uspekhi {\bf 43}, 349-364 (2000).

\bibitem{jaygerin} J.-P. Jay-Gerin, M.J. Aubin, L.G. Caron, 
Solid State Communications {\bf 21}, 771 (1977). 
\bibitem{skinner} B. Skinner, 
Phys. Rev. B {\bf 90}, 060202(R) (2014). 


\bibitem{pkbook} L. P. Pitaevskii, E. M. Lifshitz, {\it Physical
Kinetics} (Pergamon, Oxford, 1981).
\bibitem{narozhny} For e.g., unconventional MR in charge compensated systems: P. S. Alekseev, A. P. Dmitriev, I. V. Gornyi, V. Yu. Kachorovskii, B. N. Narozhny, M. Sch\"utt,and M. Titov, 
Phys. Rev. Lett. {\bf 114}, 156601 (2015), and references therein.




\bibitem{novosolev} K.S. A. Novoselov, et al. 
Nature {\bf 438}, 197-200 (2005). 

\bibitem{sonspivak} D. T. Son and B. Z. Spivak, 
Phys. Rev. B {\bf 88}, 104412 (2013).  


\bibitem{normalmetals} A. A. Abrikosov, {\it Fundamentals of the theory of metals}, North-Holland (1988).

\bibitem{prangeandgirvin} Editors: R. E. Prange, and S. M. Girvin, {\it The Quantum Hall Effect}, Springe-Verlag, New York (1987). 

\bibitem{zhao1993} See e.g., H. L. Zhao, and S.-C. Feng, 
Phys. Rev. Lett {\bf 70}, 4134 (1993).

\bibitem{renard} See e.g., V. Renard, Z. D. Kvon, G. M. Gusev, and J. C. Portal, 
Phys. Rev. B {\bf 70}, 033303 (2004).

\end{thebibliography}
\end{document}